\documentstyle[12pt,epsf]{article}
\def\PR #1 #2 #3 {Phys.~Rev.~{\bf #1}, #2 (#3)}
\def\PRL #1 #2 #3 {Phys.~Rev.~Lett.~{\bf #1}, #2 (#3)}
\def\PRD #1 #2 #3 {Phys.~Rev.~D~{\bf #1}, #2 (#3)}
\def\PLB #1 #2 #3 {Phys.~Lett.~{\bf B#1}, #2 (#3)}
\def\NPB #1 #2 #3 {Nucl.~Phys.~{\bf B#1}, #2 (#3)}
\def\RMP #1 #2 #3 {Rev.~Mod.~Phys.~{\bf #1}, #2 (#3)}
\def\ZPC #1 #2 #3 {Z.~Phys.~C~{\bf #1}, #2 (#3)}

\begin{document}
\renewcommand{\thefootnote}{\fnsymbol{footnote}}
\rightline{hep-ph/9702330} 
\medskip
\rightline{February 1997}
\bigskip\bigskip

\begin{center} {{\Large {\bf Higgs Physics: 
An Historical Perspective}\footnote{\footnotesize Presented 
at the Symposium on 
Future High Energy Colliders, Institute for Theoretical Physics,
University of California at Santa Barbara, October 21-25, 1996.}}} 
\\
\bigskip\bigskip\bigskip\bigskip
{\large\bf Scott Willenbrock} \\ 
\medskip 
Department of Physics \\
University of Illinois \\ 1110 West Green Street \\  Urbana, IL\ \ 61801 \\
\bigskip 
\end{center} 
\bigskip\bigskip\bigskip

\begin{abstract}
``Weakly-coupled'' and ``strongly-coupled'' models of electroweak symmetry
breaking are introduced by analogy with the Fermi theory of the weak 
interaction and the low-energy interaction of pions, respectively.
The implications of these two classes of models for colliders beyond the
LHC and NLC are discussed.
\end{abstract}


\setcounter{footnote}{0}
\renewcommand{\thefootnote}{\alph{footnote}}

\newpage

As we have heard repeatedly at this meeting, 
uncovering the mechanism which breaks the electroweak symmetry will be a
crucial step forward in our quest to understand nature at a deeper level.
Electroweak symmetry breaking is the target of future accelerators, and we
will hear a great deal about the phenomenology of this physics in
talks on the LHC, NLC,\footnote{NLC is used to generically denote an $e^+e^-$ 
collider with $\sqrt s = 0.5 - 1.5$ TeV.} and $\mu^+\mu^-$ colliders.  
I have therefore chosen topics with an eye towards minimizing overlap with 
other talks.  

As my title indicates, I discuss electroweak symmetry breaking from an
historical perspective.  Since we have not yet discovered the physics of
electroweak symmetry breaking, this requires some imagination.  I begin
with a very brief history of the weak interaction.  I then discuss the two
categories of electroweak symmetry breaking, usually referred to as 
``weak coupling'' and ``strong coupling.''  I next discuss the 
phenomenological implications of these two categories of electroweak 
symmetry breaking for accelerators 
beyond the LHC and NLC.  I conclude with a few historical remarks.

\section{A brief ``history'' of the weak interaction}

\indent\indent This section is intended to recall some of the milestones 
in the development 
of the standard model of the electroweak interaction.  It is too sketchy 
to be a 
proper history.  Furthermore, it is more a history of the way things could
have gone, rather than the way they actually went.  For a proper historical 
account, see Ref.~\cite{history}.

The study of the weak interaction began 100 years ago, with the discovery of
beta decay by H.~Becquerel.  It was not recognized at the time that this was 
due 
to a new force, however.  The first ``modern'' theory of the weak interaction
was due to Fermi (1933) \cite{fermi}, and involved a four-fermion 
interaction Lagrangian
\begin{equation}
{\cal L} = G_F\bar\psi\gamma^\mu\psi \bar\psi\gamma_\mu\psi
\label{fermi}
\end{equation} 
where $G_F$ is the Fermi constant and the fermion fields are those of the 
proton, neutron, electron, and electron neutrino.\footnote{Today we know 
that it is only the left-handed components of the fields that are 
involved in the interaction.  The normalization is also different from that
given in Eq.~(\ref{fermi}).}  The discovery of the 
muon in 1947 allowed the study of the weak interaction in another context, and
it was soon realized that the weak interaction is the same for electrons and 
muons.  The universality of the weak interaction suggested that it is mediated
by a gauge boson, in analogy with quantum electrodynamics, and that the 
electron and muon have the same weak ``charge.''  However, gauge bosons are 
exactly massless, while the hypothetical weak gauge boson is necessarily 
massive, since it yields the Fermi theory at energies much less than the 
weak-boson mass.  This obstacle was
surmounted in 1967 by Weinberg and Salam, who argued that the weak interaction 
is indeed a gauge theory, but with the gauge symmetry spontaneously broken,
such that the weak gauge boson acquires a mass \cite{weinbergsalam}.

We now jump to 1996, where we have a beautiful theory of the 
electroweak (and strong) interactions acting on three generations of 
quarks and leptons.
It is fair to say that the gauge interactions are understood, and that the
masses of the weak gauge bosons, the $W$ and $Z$, are understood to be a 
consequence of electroweak symmetry breaking.  However, the mechanism 
responsible for electroweak symmetry breaking is unknown.  This mechanism
is also responsible, at least in part, for the fermion masses and the 
Cabibbo-Kobayashi-Maskawa
matrix (including CP violation), so its complete elucidation is essential
to our quest to understand nature at a deeper level.

Although we do not know what the 
electroweak-symmetry-breaking mechanism is, we know that it involves new 
particles and new forces.  
It is possible that it will take another 100 years before we 
completely understand the nature of these particles and forces.

\section{Electroweak Symmetry Breaking}

\subsection{Weak coupling}

\indent\indent To explain what is meant by weakly-coupled electroweak 
symmetry breaking,
I draw an analogy with the Fermi theory, which is the low-energy limit of
a weakly-coupled theory, namely the electroweak gauge theory.

\subsubsection{Fermi theory}

\indent\indent Consider the calculation of a four-fermion amplitude in the 
Fermi theory of the
weak interaction, as shown in Fig.~1.  We use the interaction Lagrangian,
Eq.~(\ref{fermi}), to calculate the amplitude perturbatively.  We can use 
dimensional analysis to derive the dependence of the amplitude on the 
typical energy in the process, $E$. The Fermi 
constant, $G_F$, has units of inverse energy squared.  Since the amplitude is
dimensionless, the leading term in the amplitude, from the diagram in 
Fig.~1(a), is proportional to $G_FE^2$.

\begin{figure}[tb]
\begin{center}
\epsfxsize= \textwidth   
\leavevmode
\epsfbox{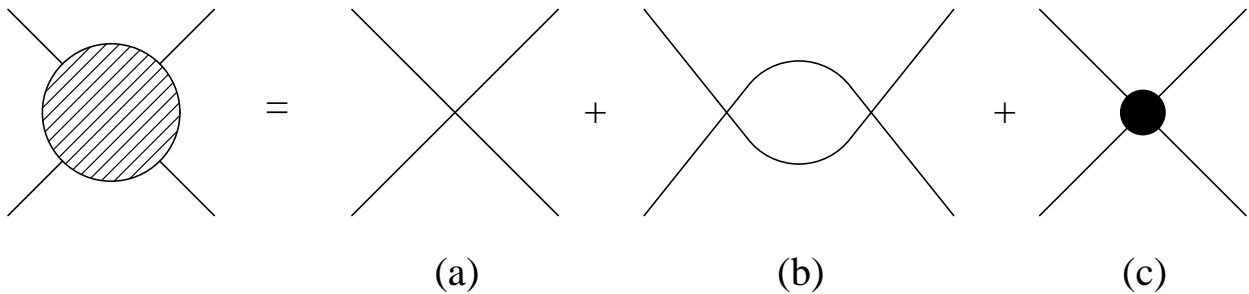}
\end{center}
\caption[fake]{Fermion scattering in the Fermi
theory of the weak interaction: (a) leading order, (b) next-to-leading-order
correction, with the loop-momentum integration cut off at $\Lambda$, 
(c) next-to-leading-order correction from loop momenta greater than $\Lambda$.}
\end{figure}

At next order in perturbation theory, one has the one-loop diagram in 
Fig.~1(b).
Using dimensional analysis, we see that this diagram is proportional to 
$G_F^2E^4$.  Thus we are performing a perturbative expansion in
powers of the dimensionless quantity $G_FE^2$.

This statement is in fact correct, but the argument is somewhat more subtle,
because the loop integration is ultraviolet divergent.  The 
modern attitude towards this divergence is encompassed by the idea of an
``effective field theory'' \cite{weinberg,weinbergqft}.  
Let's say we only believe the Fermi theory up
to some energy $\Lambda>>E$.  We restrict the loop integration to 
momenta less 
than $\Lambda$, and profess ignorance about what happens for momenta greater 
than $\Lambda$.  However, we cannot simply throw away the contribution from
momenta greater than $\Lambda$.\footnote{This is reminiscent of a quote from
the early days of renormalization theory: ``Just because something is infinite
does not mean it is zero!''\cite{quote}.}  Instead, we parameterize it
by adding additional terms to the interaction Lagrangian of the form
\begin{equation}
{\cal L} = c\,G_F^2\partial^\nu(\bar\psi\gamma^\mu\psi) 
\partial_\nu (\bar\psi\gamma_\mu\psi) + ...
\label{effective}
\end{equation} 
which are characterized by having two derivatives, and an unknown 
dimensionless coefficient
$c$. These terms yield a tree-level contribution
to the amplitude proportional to $c\,G_F^2E^4$, indicated in Fig.~1(c).  
Thus the next-to-leading-order amplitude is given schematically by
\begin{equation}
A = G_FE^2 + G_F^2E^4 \ln\Lambda + c\,G_F^2E^4
\end{equation} 
where the ultraviolet divergence of the loop diagram is evidenced by 
the dependence of the amplitude on $\ln\Lambda$.
Combining the last two terms using $c^{\prime} \equiv c + \ln\Lambda$ yields
\begin{equation}
A = G_FE^2 + c^{\prime}\,G_F^2E^4
\end{equation} 
where $c^{\prime}$ is to be taken from experiment.  Thus the amplitude is
indeed an expansion in powers of the dimensionless quantity $G_FE^2$, despite
the ultraviolet divergence.

At low energies, $E<<G_F^{-1/2}$, only the first few terms in the expansion
are numerically important, and the amplitude depends on just a few 
coefficients which must be extracted from experiment.  However, for 
$E\sim G_F^{-1/2}$, every term in the expansion is equally important, and 
the amplitude depends on an infinite number of unknown coefficients.  The 
theory therefore loses all predictive power.  It is sometimes said that the
theory ``breaks down'' or becomes ``strongly-interacting,'' but in fact the 
theory simply becomes useless.

There is a quick way to estimate the critical energy at which the theory
becomes unpredictive, which also allows us to get the numerical factors
straight.  Unitarity implies that the partial waves of a two-particle 
scattering amplitude
have a real part which does not exceed 1/2 in magnitude.  If the leading
term in the expansion were to greatly exceed this bound, the higher-order
terms
in the expansion would have to be just as large as the leading term in order
to restore unitarity.  But if that were the case, then every term in the 
expansion is equally important, and the theory is unpredictive.  
So we can estimate the critical
energy by imposing the unitarity bound on the leading-order amplitude.  
This yields \cite{unitarity}
\begin{equation}
E_{critical}\approx \left(\frac{\sqrt 2 \pi}{G_F}\right)^{1/2}
\approx 600 \;{\rm GeV}.
\label{ecfermi}
\end{equation}
Thus the Fermi theory loses predictive power for $E\sim 600$ GeV.

It is a logical possibility that physics simply becomes very complicated 
at energies in excess of the critical energy.  However, history has taught
us to expect the opposite: the theory becomes simpler at higher energy.
In the case of the Fermi theory, the four-fermion interaction is replaced
by the exchange of a weak gauge boson, as shown in Fig.~2.  The amplitude
is proportional to 
\begin{equation}
A \sim g^2\frac{E^2}{E^2-M_W^2}
\end{equation} 
where $g$ is the gauge coupling, and the denominator is from the $W$ 
propagator.  Since 
\begin{equation}
G_F \sim g^2/M_W^2
\label{gf}
\end{equation}
the amplitude reduces to that of 
the Fermi
theory for $E<<M_W$.  However, for $E>>M_W$, the amplitude is proportional to
$g^2$.  Hence the expansion parameter is a (small) constant, 
and the theory is predictive for all energies.

\begin{figure}[tb]
\begin{center}
\epsfxsize= 0.64\textwidth   
\leavevmode
\epsfbox{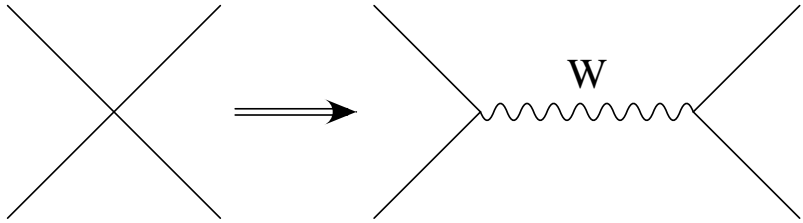}
\end{center}
\caption[fake]{The Fermi theory of the weak interaction is the low-energy
approximation to a gauge theory.}
\end{figure}

In the case of the Fermi theory, new physics, in the form of the $W$ boson, 
enters before the critical energy. Formally,
\begin{equation}
M_W^2 \sim \frac{g^2}{4\pi} E_{critical}^2 < E_{critical}^2\;,
\end{equation}
using Eqs.~(\ref{ecfermi}) and (\ref{gf}).\footnote{More precisely, using
$G_F/\sqrt 2=g^2/8M_W^2$.}
The fact that the new physics enters before the critical energy can thus be 
related to the fact that $g^2/4\pi<1$, i.e., that the theory is weakly coupled.

\subsubsection{Electroweak theory}

\indent\indent In the previous section, the $W$ boson was responsible for 
regulating the 
growth of the four-fermion amplitude at high energy.  Now consider the 
scattering of the $W$ bosons themselves, in particular 
longitudinal (helicity-zero) $W$ bosons, as shown in Fig.~3.
The top row of Feynman diagrams depicts the gauge interactions responsible 
for the scattering.  The resulting amplitude is proportional to $G_FE^2$,
just as in the case of the Fermi theory.  Using unitarity, we estimate the
critical energy to be \cite{chanowitz}
\begin{equation}
E_{critical}\approx \left(\frac{4\sqrt 2 \pi}{G_F}\right)^{1/2} 
\approx 1.2\;{\rm TeV}\;.
\end{equation}
Thus the electroweak theory loses predictive power for $E\sim 1.2$ TeV.

\begin{figure}[tb]
\begin{center}
\epsfxsize= \textwidth   
\leavevmode
\epsfbox{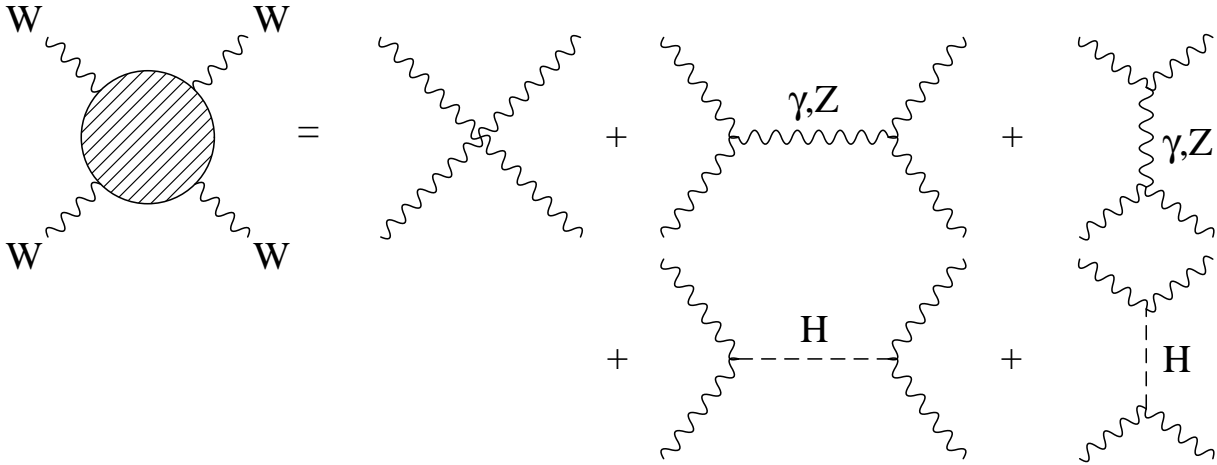}
\end{center}
\caption[fake]{Weak-vector-boson scattering in the standard model of the
electroweak interaction.}
\end{figure}

By analogy with the Fermi theory, we might expect new physics, in the form of 
a new
particle, to regulate the growth of the amplitude.  This is exactly what 
happens in the standard Higgs model.  The Higgs boson gives rise to 
additional Feynman diagrams, shown in the second row in Fig.~3.  These 
diagrams cancel the terms proportional to $G_FE^2$, leaving behind an 
amplitude proportional to $G_Fm_H^2 \sim \lambda$,
where $\lambda$ is the Higgs self-coupling.
Thus the expansion parameter is a (small) constant, and the theory is 
predictive for
all energies.\footnote{Ignoring the running of $\lambda$, to be discussed 
in a later footnote.}  
Higgs-higgs scattering is also proportional to $\lambda$, so 
the theory is complete.

The formal relation between the Higgs mass and the critical energy is
\begin{equation}
m_H^2 \sim \frac{\lambda}{4\pi} E_{critical}^2 < E_{critical}^2
\end{equation}
where the last relation relies on $\lambda/4\pi < 1$, 
known from non-perturbative 
studies of the standard Higgs model \cite{luscher}.  Thus the new physics,
namely the Higgs boson, enters before the critical energy because the theory 
is weakly coupled.  In fact, it has been shown that the standard Higgs model,
and variations of it (such as two Higgs doublets, as used in 
supersymmetric models), are the only weakly-coupled
models of electroweak symmetry breaking \cite{cornwall}. 

\subsection{Strong coupling}

\indent\indent To explain what is meant by strongly-coupled electroweak 
symmetry breaking,
I draw an analogy with low-energy pion physics, which is the low-energy limit 
of a strongly-coupled theory, namely QCD.

\subsubsection{Pion theory}

\indent\indent The low-energy interaction of pions is dictated by the 
fact that they are
the (approximate) Goldstone bosons of spontaneously-broken chiral symmetry.
This is embodied by an effective field theory of pions called 
``chiral perturbation theory'' \cite{weinberg}.
The leading interaction of pions is
\begin{equation}
{\cal L}=G_\pi {\bf \pi}\cdot\partial^\mu{\bf \pi} 
{\bf \pi}\cdot\partial_\mu{\bf \pi}
\end{equation}
where $G_\pi =1/(2f_\pi^2)$, with $f_\pi$ the pion decay constant. The coupling
$G_\pi$ has dimensions of inverse energy squared, in analogy with the Fermi
constant, $G_F$.

The leading-order amplitude for $\pi\pi$ scattering is proportional to 
$G_\pi E^2$.  Using unitarity to estimate the critical energy at which the 
theory loses predictive power yields
\begin{equation}
E_{critical}\sim \left(\frac{4\pi}{G_\pi}\right)^{1/2} 
\approx 450 \;{\rm MeV} \;.
\end{equation}

What actually happens in nature in $\pi\pi$ scattering at 450 MeV?  One 
begins to encounter a plethora of new particles, beginning with the 
$\sigma$ meson.\footnote{After a twenty-year absence, the $\sigma$ meson has 
once again been recognized by the  
Particle Data Group, but they are hedging on its mass: $f_0(400-1400)$
\cite{pdg}.} 
The theory of pion interactions becomes complicated above 450 MeV, and loses
all predictive power.  Unlike the case of a weakly-coupled theory, there
is no new particle (the analogue of a $W$ or a Higgs boson) that 
restores predictivity to the theory.\footnote{One might be tempted to interpret
the $\sigma$ as the Higgs boson of the pion theory.  This interpretation 
is invalidated by the $\rho(770)$, which is comparable in mass to the $\sigma$.
A true Higgs theory would have no such particle.}

Although the pion theory becomes complicated above 450 MeV, we have learned
that nature is nevertheless simple: there is a new description of physics 
in terms of quarks and gluons, interacting via QCD.  At low energies 
this theory is strongly-interacting, and gives rise to all the complications
of hadron physics.  But even at low energies the theory itself is simple,
in the sense that it is described by a Lagrangian with just a few terms in
it, and the only parameters are the strong coupling constant 
and the quark masses.

\subsubsection{Electroweak theory}

\indent\indent Now consider $WW$ scattering near the critical energy of 
1.2 TeV, and imagine
that nature does not provide a Higgs boson to regulate the growth of the 
amplitude with energy.  What will happen?  There is no answer to this 
question within the electroweak theory; it simply becomes unpredictive.
However, based on our experience with pion physics, we might expect that
there is a deeper, simpler theory, which is hidden from view by virtue
of the fact that it is strongly interacting, and therefore manifests itself
in a complicated way.  This would mean that there are new particles in nature,
experiencing a new strong interaction.

For example, one can imagine that this simpler theory is analogous 
to QCD.  This is the idea behind the so-called Technicolor theory 
\cite{weinbergtc,susskind}.  
In that case, one would encounter the analogues of the
$\sigma$, $\rho(770)$, etc., at energies above 1.2 TeV.  These 
``Technimesons'' would be made from strongly-interacting ``Techniquarks,'' 
bound together by ``Technigluons.''  This class of models is reviewed
by Appelquist at this symposium.

\section{Phenomenology beyond the LHC/NLC}

\indent\indent In this section I make some observations about the 
phenomenological 
implications of electroweak symmetry breaking for colliders with energy in 
excess of the LHC ($\sqrt s = 14$ TeV) and the NLC ($\sqrt s = 0.5 - 1.5$ 
TeV).  I begin where I left off, with the case of strongly-coupled 
electroweak symmetry breaking.

\subsection{Strong coupling}

\indent\indent Let's continue our analogy with pion physics.  To study the 
strong 
interaction at low energy, one performs scattering experiments involving 
pions.  For example, one can perform $\pi\pi$ scattering experiments, or 
study the coupling of the pion to virtual photons, as depicted in Fig.~4(a).  
The analogues of these processes for the electroweak interaction are  
shown in Fig.~4(b), where the longitudinal $W$ bosons replace the pions.  
The incident fermions could be either quarks and antiquarks (LHC) or 
electrons and positrons (NLC).  These colliders will be capable of probing
these processes at energies of about 1 TeV, comparable to the critical energy
of 1.2 TeV at which the theory loses predictive power.

\begin{figure}[tb]
\begin{center}
\epsfxsize= 0.9\textwidth   
\leavevmode
\epsfbox{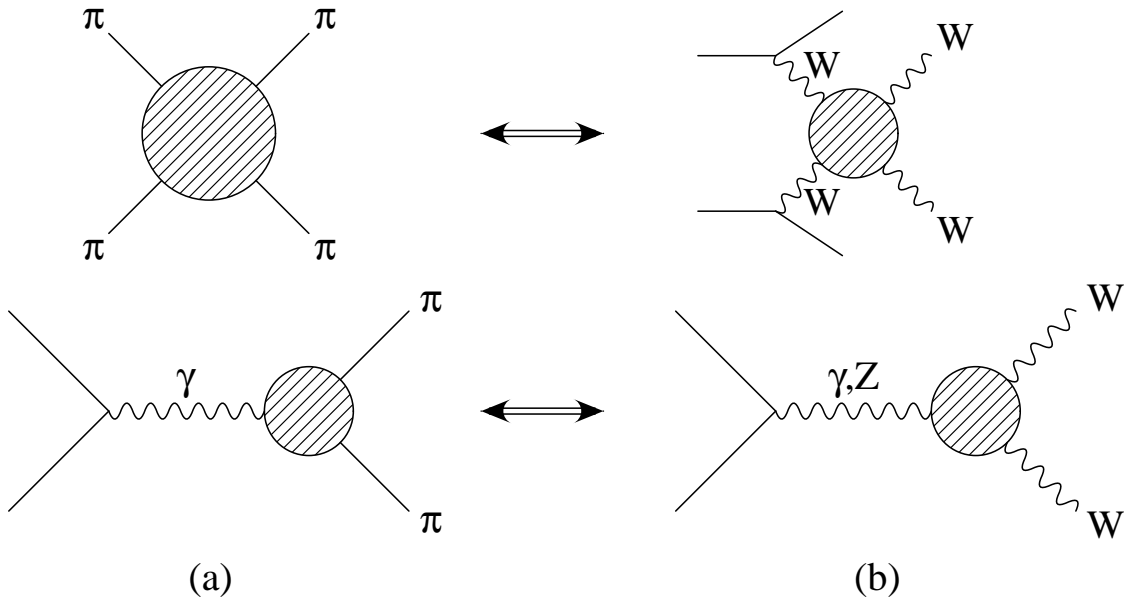}
\end{center}
\caption[fake]{(a) Experimental probes of the pion and 
(b) their counterparts for weak vector bosons.}
\end{figure}

Recall that the critical energy for $\pi\pi$ scattering is about 450 MeV.  How 
much would we know about the strong interaction if we had data from
$\pi\pi$ scattering and $\gamma^*\to \pi\pi$ at energies only up to 450 MeV?  
The answer is very little.  We might know about the $\sigma$, but we would
see only the low-energy tail of the $\rho$, and the heavier mesons would
be completely out of sight.  More importantly, we would not know that the 
mesons are composed of strongly-interacting quarks, interacting via QCD.

The moral is that if the mechanism of electroweak symmetry breaking is 
indeed strongly-coupled, it will likely require energies greatly in 
excess of 1 TeV for the complete elucidation of this physics.  This 
implies the need for colliders beyond the LHC and NLC.  Although these 
machines are likely to tell us {\em something} about strongly-coupled 
electroweak
symmetry breaking, it is hard to imagine they will be able to tell us
{\em everything} about it.

\subsection{Weak coupling}

\indent\indent The case of weakly-coupled electroweak symmetry breaking is 
apparently in 
stark contrast with that of the strongly-coupled case.   For example, 
the standard Higgs
model has a Higgs boson with mass less than 700 GeV \cite{luscher}, 
and nothing else.  
The LHC and NLC are both capable of discovering this particle.  Once 
discovered, and its coupling to itself and other particles measured, we 
have learned everything there is to learn about electroweak symmetry 
breaking.  Can it really be this simple?

Probably not.  Although the standard Higgs model is predictive at all 
energies, it suffers from another disease - it is ``unnatural'' 
\cite{'t Hooft}.  We know 
that the Higgs field acquires a vacuum-expectation value of 
$v=(\sqrt 2 G_F)^{-1/2}\approx 250$ GeV.  The diagram in Fig.~5(a) 
is one of the one-loop corrections to the vacuum-expectation value, from a 
Higgs loop.  There are similar one-loop corrections from loops of weak 
bosons and fermions.  All these one-loop diagrams share the
feature that they are quadratically divergent \cite{susskind}.  
Let us regard the standard 
Higgs model as being valid up to some energy $\Lambda$.  The relation 
between the bare vacuum-expectation value, $v_0$, and the actual 
vacuum-expectation value, $v$, may be approximated by cutting off the loop
integration at momenta of order $\Lambda$:
\begin{equation}
v^2=v_0^2+O\left(\frac{\Lambda^2}{(4\pi)^2}\right)
\end{equation}
where the factor $(4\pi)^2$ is the usual factor which arises from loop 
diagrams.  There is also an (unknown) contribution from momenta 
greater than $\Lambda$, but we will assume that it does not conspire to 
cancel the contribution from momenta less than $\Lambda$.\footnote{It has 
been argued that just such a cancellation occurs if the underlying theory, 
to which the standard Higgs model is a low-energy approximation, 
is conformally invariant \cite{bardeen}. However, a realistic example 
of such a theory has yet to be constructed.}
If $\Lambda >> v$, then $v_0$ 
must be tuned to almost exactly cancel the one-loop contribution to 
the vacuum-expectation value.  Such a fine tuning is unnatural.
Instead, it is natural to expect 
$\Lambda/(4\pi) \leq v$.  Thus if nature makes use of the standard Higgs 
model, we anticipate that it is replaced by a more fundamental theory at 
energy $\Lambda \leq 4\pi v \approx 3$ TeV, regardless of the Higgs 
mass.\footnote{A separate argument can 
be given for the incompleteness of the standard Higgs model.  The Higgs 
self-coupling, $\lambda$, is a running coupling, and increases
with energy (for sufficiently-large Higgs mass), eventually blowing up.  New 
physics must intervene before this occurs. This can be used to place an upper
bound on the Higgs mass for a given energy scale of the new physics
\cite{lindner}. For example, if the new physics 
is at the Planck scale, there is an upper bound on the Higgs mass of 
about 200 GeV \cite{maini}. 
However, naturalness implies that there is new physics by at least
3 TeV, so it is not realistic to imagine that there is no new physics 
until the Planck scale.  Imposing the condition that the Higgs coupling not
blow up below 3 TeV yields an upper bound on the Higgs mass of only about 
700 GeV.}

\begin{figure}[tb]
\begin{center}
\epsfxsize= 0.725\textwidth   
\leavevmode
\epsfbox{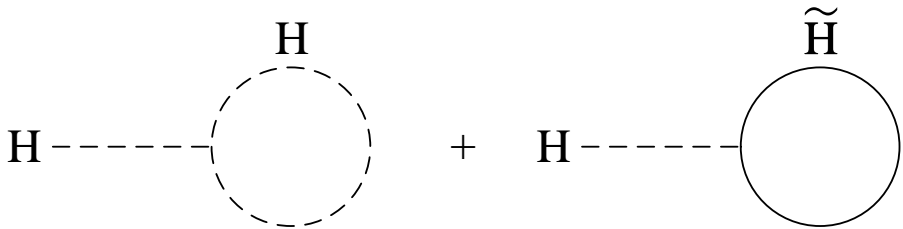}
\end{center}
\caption[fake]{(a) Quadratically-divergent one-loop correction to the Higgs
vacuum-expectation value from a Higgs loop; (b) the quadratic 
divergence is cancelled by a Higgsino loop in a supersymmetric theory.}
\end{figure}

Thus we see that weakly-coupled electroweak symmetry breaking may not be 
so different from strongly-coupled electroweak symmetry breaking.  The 
strongly-coupled approach involves new physics at the TeV scale, while the 
weakly-coupled approach suggests new physics (beyond the standard Higgs model)
by at least the TeV scale.  It 
could be that this new physics lies well below the TeV scale, in which case
it may be possible to discover all of it with the LHC/NLC.  But if it 
really lies at the TeV scale, it will likely require colliders beyond the 
LHC/NLC for its complete elucidation.

A well-known example of new physics in the weakly-coupled scenario is 
supersymmetry.  If nature is supersymmetric, every particle is 
accompanied by a superpartner.  The one-loop correction to the Higgs 
vacuum-expectation value in Fig.~5(a) is accompanied by a second diagram,  
shown in Fig.~5(b), in which the Higgs loop is replaced by a Higgsino loop.  
The quadratic divergences of each diagram cancel, and the theory becomes 
natural.  The same sort of cancellation occurs for loops of weak bosons 
and fermions and their superpartners.  

If nature is supersymmetric, then supersymmetry must be broken, since we 
don't observe the superpartners of the known particles.  Let's refer to 
the mass scale of the superpartners as $M_{SUSY}$.  It $M_{SUSY}$ were much
larger than $v$, the quadratic divergence would reappear, cut off only
when the momenta reach $M_{SUSY}$:
\begin{equation}
v^2=v_0^2+O\left(\frac{M_{SUSY}^2}{(4\pi)^2}\right)\;.
\end{equation}
Thus the model is only natural if $M_{SUSY}<4\pi v \approx$ 3 TeV.  More 
careful estimates yield a somewhat lower scale \cite{anderson}.

If $M_{SUSY}$ is at the TeV scale, it may require colliders beyond the 
LHC/NLC to discover all the superpartners and measure 
their couplings.
However, if $M_{SUSY}$ is a few hundred GeV, then all the physics of 
supersymmetry may be within reach of the LHC/NLC.  This is the only natural 
scenario I know of in which the entire physics of electroweak symmetry 
breaking is elucidated by these colliders.  However, even in this scenario, 
there may 
be motivation for higher-energy colliders.  The physics of supersymmetry 
breaking may lie at the 10 TeV scale, as in the class of models 
in which dynamical supersymmetry breaking is communicated to the standard
model 
via new gauge 
interactions \cite{dine}.  It is possible that the elucidation of the 
physics of electroweak symmetry breaking will reveal to us the scale of 
supersymmetry breaking.

It is also possible that the new physics which subsumes the standard Higgs
model at the TeV scale is strongly-interacting, and has nothing to do with
supersymmetry \cite{kaplan}.  This class of models is similar to the 
strongly-coupled electroweak-symmetry-breaking scenario, with the 
exception that there is a Higgs boson.  The complete elucidation of these
models would likely require colliders beyond the LHC/NLC, as in the 
strongly-coupled models.

\section{Concluding historical remarks}

\indent\indent The history of the weak interaction was punctuated by 
periods of confusion, 
followed by clarification, which ultimately led to a beautiful theory.
For example, the early measurements of beta decay indicated that the 
emitted electron was monoenergetic.  It took twenty years to establish that 
the electron is emitted with a continuous spectrum of 
energies.  This was the first evidence for the existence of the 
neutrino, which was a vital ingredient in Fermi's theory of the weak 
interaction \cite{history1}.

The same sequence of events is likely to occur for the 
electroweak-symmetry-breaking mechanism.  In fact, we are already in the 
first stage, the 
period of confusion.  There are already several experimental results which 
have been interpreted as a Higgs boson 
\cite{mckay,ma,willey,georgi,glashow,haber,shin}.  
These interpretations have been made either in the context of a multi-Higgs 
model, or with additional new physics -- none has had an interpretation in 
terms of the standard Higgs model with no new physics.  

A particularly 
noteworthy example is Ref.~\cite{glashow}, which was an attempt to 
interpret the $\zeta(8.3)$ as a Higgs boson with enhanced coupling to $b$ 
quarks.\footnote{The $\zeta(8.3)$ was conjectured to be responsible for 
a monoenergetic photon signal in $\Upsilon$ decay, via
$\Upsilon\to\zeta\gamma$.  The experiment turned out to be erroneous.}
This would imply an enhanced coupling to muons, which led the 
authors to conclude that if their interpretation were correct, ``the case
for construction of a dedicated muon collider for Higgs boson studies 
will become as compelling as it is technically feasible.''  The idea of 
using a muon collider for resonant Higgs production (even the standard Higgs) 
has lately resurfaced and been considerably refined \cite{barger}. 

My final historical remark concerns our ability to foresee the physics of 
the future.  There are many examples of incorrect theories and 
prognostications, many more than correct ones, and these are often held 
up as examples of our inability to predict what will be found in the next
generation of colliders.  While I have some sympathy for this point, 
I also feel it can be overstated.  A well-known example is the top-quark 
mass; predictions ranged from 15 GeV (just above the reach of PEP) to 230 
GeV, and everything in between.  But perhaps this makes us lose sight of 
the real achievement: we knew the top quark existed long before it was
discovered, based entirely on indirect evidence.

Today many feel that there is indirect evidence for weak-scale 
supersymmetry.  Although the evidence is not as compelling as it was for
the top quark, an advocate might make the following statement: 
\begin{quote}
``There can be no two opinions about the practical utility of 
supersymmetry, but until clear experimental evidence for the existence of 
supersymmetry can be obtained, supersymmetry must remain purely 
hypothetical.  Failure to detect any evidence of supersymmetry is no 
evidence against its existence.''
\end{quote}
In fact, this quote is paraphrased from a 1937 meeting of the Royal 
Society, except I have substituted ``supersymmetry'' for ``the neutrino''
throughout \cite{history2}.  

\section*{Acknowledgements}

\indent\indent I am grateful for conversations with D.~Dicus, R.~Leigh,
and W.~Marciano, and for assistance from Z.~Sullivan.
This work was supported in part by the 
Department of Energy under Grant No.~DE-FG02-91ER40677 and by the 
National Science Foundation under Grant No.~PHY94-07194. 

 \clearpage

\vfill


\begin{thebibliography}{99}

\bibitem{history} A.~Pais, {\sl Inward Bound} (Oxford University Press, Oxford,
1986).

\bibitem{fermi} E.~Fermi, Z.~Phys.~{\bf 88} 161 (1934).

\bibitem{weinbergsalam} S.~Weinberg, \PRL 19 1264 1967 ; A.~Salam, in 
{\sl Elementary Particle Physics}, edited by N.~Svartholm 
(Almqvist and Wiksell, Stockholm, 1968), p.~367.

\bibitem{weinberg} S.~Weinberg, Physica {\bf 96A}, 327 (1979).

\bibitem{weinbergqft} S.~Weinberg, {\sl The Quantum Theory of Fields}
(Cambridge University Press, Cambridge, 1995), Vol.~I.

\bibitem{quote} Ref.~\cite{weinbergqft}, p.~35.

\bibitem{unitarity} E.~Abers and B.~Lee, Phys.~Rep.~{\bf 9}, 1 (1973).

\bibitem{chanowitz} M.~Chanowitz and M.~K.~Gaillard, \NPB 261 379 1985 ;
W.~Marciano, G.~Valencia, and S.~Willenbrock, \PRD 40 1725 1989 .

\bibitem{luscher} M.~L\"uscher and P.~Weisz, \PLB 212 472 1988 .

\bibitem{cornwall} J.~M.~Cornwall, D.~Levin, and G.~Tiktopoulos, 
\PRL 30 1268 1973 ; \PRD 10 1145 1974 ; C.~Llewellyn-Smith, \PLB 46 233 1973 .

\bibitem{pdg} Review of Particle Properties, \PRD 54 1 1996 .

\bibitem{weinbergtc} S.~Weinberg, \PRD 19 1277 1979 .

\bibitem{susskind} L.~Susskind, \PRD 20 2619 1979 .

\bibitem{'t Hooft} G.~'t Hooft, in {\sl Recent Developments in Gauge Theories},
Proceedings of the NATO Advanced Study Institute, Carg\`ese, 1979,
eds.~G.~'t Hooft {\it et al.} (Plenum, New York, 1980), p.~135. 

\bibitem{bardeen} W.~Bardeen, FERMILAB-CONF-95-391-T, to appear in the 
{\sl Proceedings of the 1995 Ontake Summer Institute}, Ontake Mountain, 
Japan, Aug.~27 - Sept.~2, 1995.

\bibitem{lindner} M.~Lindner, \ZPC 31 295 1986 .

\bibitem{maini} L.~Maini, G.~Parisi, and R.~Petronzio, \NPB 136 115 1978 ;
N.~Cabbibo, L.~Maini, G.~Parisi, and R.~Petronzio, \NPB 158 295 1979 .

\bibitem{anderson} G.~Anderson and D.~Castano, \PRD 52 1693 1995 .

\bibitem{dine} M.~Dine and A.~Nelson, \PRD 48 1277 1993 .

\bibitem{kaplan} D.~Kaplan and H.~Georgi, \PLB 136 183 1984 ;
D.~Kaplan, S.~Dimopoulos, and H.~Georgi, \PLB 136 187 1984 .

\bibitem{history1} Ref.~\cite{history}, p.~11.

\bibitem{mckay} D.~McKay and H.~Munczek, \PRL 34 432 1975 .

\bibitem{ma} E.~Ma, S.~Pakvasa, and S.~Tuan, \PRD 16 568 1977 .

\bibitem{willey} R.~Willey, \PRL 52 585 1984 .

\bibitem{georgi} H.~Georgi and S.~Glashow, \PLB 143 155 1984 .

\bibitem{glashow} S.~Glashow and M.~Machacek, \PLB 145 302 1984 .

\bibitem{haber} H.~Haber and G.~Kane, \NPB 250 716 1985 .

\bibitem{shin} M.~Shin, H.~Georgi, and M.~Axenides, \NPB 253 205 1985 .

\bibitem{barger} V.~Barger, M.~Berger, J.~Gunion, and T.~Han, \PRL 75 1462 
1995 ; hep-ph/9602415.

\bibitem{history2} Ref.~\cite{history}, p.~18.

\end{thebibliography}
\end{document}